\newlength{\absize}
\documentclass[12pt]{article}
\usepackage{graphicx}
\usepackage{latexsym}
\usepackage{amssymb}
\usepackage{bbm}
\usepackage{setspace}
\usepackage{braket}
\usepackage{bm}
\renewcommand{\arraystretch}{2}

\input prepictex
\input pictex
\input postpictex
\newdimen\tdim
\tdim=\unitlength
\def\stpltsmbl{\setplotsymbol ({\small .})}

\newbox\sru
\setbox\sru=\hbox{\beginpicture
\setcoordinatesystem units <\tdim,\tdim>
\stpltsmbl
\setquadratic
\plot
  0.0   0.0
  4.8   1.5
  7.5   5.0
  7.3   8.5
  5.0  10.0
  2.7   8.5
  2.5   5.0
  5.2   1.5
 10.0   0.0
/
\endpicture}
\def\springru #1 #2 *#3 /{\multiput {\copy\sru}  at
#1 #2 *#3 10 0 /}

\renewcommand{\bar}{\overline}

\newcommand{\spur}[1]{\!\not\! #1 \,}

\newcommand{\cL}{\mathcal{L}}

\newcommand{\cO}{\mathcal{O}}

\newcommand{\cB}{\mathcal{B}}
\newcommand{\cC}{\mathcal{C}}

\newcommand{\pd}{\partial}

\renewcommand{\slash}[1]{#1\!\!\!/}

\newcommand{\be}{\begin{equation}}
\newcommand{\ee}{\end{equation}}
\newcommand{\bea}{\begin{eqnarray}}
\newcommand{\eea}{\end{eqnarray}}

\newcommand{\comment}[1]{}

\newcommand{\cc}{{\displaystyle*}}

\begin{document}

\thispagestyle{empty}
\pagestyle{empty}
\newcommand{\starttext}{\newpage\normalsize
 \pagestyle{plain}
 \setlength{\baselineskip}{3ex}\par
 \setcounter{footnote}{0}
 \renewcommand{\thefootnote}{\arabic{footnote}}
 }
\newcommand{\preprint}[1]{\begin{flushright}
 \setlength{\baselineskip}{3ex}#1\end{flushright}}
\renewcommand{\title}[1]{\begin{center}\LARGE
 #1\end{center}\par}
\renewcommand{\author}[1]{\vspace{2ex}{\large\begin{center}
 \setlength{\baselineskip}{3ex}#1\par\end{center}}}
\renewcommand{\thanks}[1]{\footnote{#1}}
\renewcommand{\abstract}[1]{\vspace{2ex}\normalsize\begin{center}
 \centerline{\bf Abstract}\par\vspace{2ex}\parbox{\absize}{#1
 \setlength{\baselineskip}{2.5ex}\par}
 \end{center}}

\title{Automatic Fine-Tuning in the 2-Flavor Schwinger Model}
\author{
 Howard~Georgi\thanks{\noindent \tt hgeorgi@fas.harvard.edu}\\
Center for the Fundamental Laws of Nature\\
Jefferson Physical Laboratory \\
Harvard University \\
Cambridge, MA 02138\\
}
\date{\today}
\abstract{
I discuss the 2-flavor Schwinger model both without and with fermion
masses.  I argue that the phenomenon of ``conformal coalescence,'' in
unparticle physics in which linear combinations of short distance operators
can disappear from the long-distance theory, makes it easy to understand
some puzzling features of the model with small fermion masses.  In
particular, I argue that for an average fermion mass $m_f$ and a mass
difference $\delta m$, so long as both are small compared to the dynamical
gauge boson mass $m=e\sqrt{2/\pi}$, isospin breaking effects in the low
energy theory are {\bf exponentially suppressed by powers of
\boldmath{$\exp\Bigl(-(m/m_f)^{2/3}\Bigr)$} even if
\boldmath{$\delta m\approx m_f$}!}  In the low energy theory, this looks
like exponential fine-tuning, but it is done automatically by conformal
coalescence.

}

\starttext
\section{The Schwinger Model with 2 Flavors\label{sec-2-flavor}}

In this note, I am in the uncomfortable position of (mildly) disagreeing
with one of my intellectual heroes, the great Sidney Coleman who (alas) is
no longer around to correct me.  In his classic paper, 
``More about the massive Schwinger
model''~\cite{Coleman:1976uz} (which mostly concerns the
model without 
flavor) Coleman briefly discusses the 2-flavor model and 
identifies three puzzles.

{\it In the one-quark theory, everything that happened, even for strong coupling,
was qualitatively understandable in terms of the basic ideas of the naive quark
model, the picture of quarks confined in a linear potential. For the two-quark
theory, there are three strong-coupling phenomena that I cannot understand in
these terms:\\
(1) Why are the lightest particles in the theory a degenerate isotriplet, even if
one quark is 10 times heavier than the other?\\
(2) Why does the next-lightest particle have $I^{PG}= 0^{++}$, rather than $0^{--}$?\\
(3) For $|\theta|=\pi$, how can an isodoublet quark and an isodoublet antiquark,
carrying opposite electric charges, make an isodoublet bound state with electric
charge zero?} 

I will argue that by taking proper account of the unparticle physics
of the massless model we can easily resolve the first two and understand
why the third is not puzzling.\footnote{The ideas in this paper are closely related to
the analysis of diagonal color models in 1+1~\cite{Georgi:2019tch}.
See also
\cite{Belvedere:1978fj, GamboaSaravi:1981zd, Gattringer:1993ec,
Delphenich:1997ex}.  Two papers that I know of - \cite{Nagy:2008ys} and
\cite{Azcoiti:2019moz} - address Coleman's puzzles explicitly, but I think that their
suggestions are quite different from mine.}   I will
suggest that the resolution of the first puzzle is a new mechanism for
exponential fine-tuning. 

The 
Lagrangian is
\be
\cL=
\left( \sum_{j=1}^{2}\bar\psi_j\,(i\spur\pd - e\slash A)\,\psi_j\right)
-\frac{1}{4}F^{\mu\nu} F_{\mu\nu}
-m_f\,\sum_{j=1}^{2}\bar\psi_j\psi_j
\label{2f}
\ee
I begin by discussing $m_f=0$ and consider the mass term in
section~\ref{sec-mass}.   

For gauge invariant correlators of 
local fields, the result of summing the perturbation theory to all orders
can be found simply by making the following replacements:
\be
A^\mu = \epsilon^{\mu\nu}\pd_\nu(\cB-\cC)/m
\label{f-A-decomposition}
\ee
\be
\psi_j = e^{-i(\pi/2)^{1/2}\,(\cC-\cB)\gamma^5}\Psi_j
\label{f-psi-redef-vcb}
\ee
where
\begin{equation}
m^2=2e^2/\pi
\label{f-m-m0}
\end{equation}
with the free-field Lagrangian
\be
\cL_f=\left(\sum_{j=1}^{2}i\bar\Psi_j\spur\pd\Psi_j\right)
-\frac{m^2}{2}\cB^2+\frac{1}{2}\pd_\mu\cB\pd^\mu\cB
- \frac{1}{2}\pd_\mu\cC\pd^\mu\cC\\
\label{f-Sommerfield-redefined2}
\ee
So that $\Psi_j$ for $j=1$ to $2$ is a free doublet of fermion fields, $\cB$ is a
spinless field with mass $m$,
and $\cC$ is a massless ghost.

The massless model has a classical chiral $U(2)\times U(2)$ symmetry
broken by the chiral anomaly down to $SU(2)\times SU(2)\times U(1)$.  
It is a Banks-Zaks model~\cite{Banks:1981nn} with free-fermion behavior at
distances much smaller than $1/m$ and a low-energy sector with conformal
symmetry for distances much larger than $1/m$.\footnote{I will sometimes
follow Coleman's lead and refer to the fermions as ``quarks.''}
There are fermion-anti-fermion operators
transforming
like the $(2,2)$ representation of the chiral $SU(2)\times SU(2)$ symmetry.
They are shown 
below
\begin{equation}
O^j_{k12}=\psi^*_{j\,1}\psi_{k\,2}
\mbox{~~and~~}
O^j_{k21}=\psi^*_{j\,2}\psi_{k\,1}
\label{s2d1/2}
\end{equation}
Where $j$ and $k$ are flavor indices, $\gamma^5\psi_{k\,1}=\psi_{k\,1}$,
and $\gamma^5\psi_{k\,2}=-\psi_{k\,2}$
To infinite order in perturbation theory, these flow at long distances to
independent unparticle 
operators with dimension $1/2$. 
But this changes due to
non-perturbative effects associated with the 
chiral
$SU(2)\times SU(2)$ singlet operators\footnote{For later convenience, we
have reversed the order of the flavors on the $\psi$s compared to the $\psi^*$s.}  
\begin{equation}
O^{z}_{12}\equiv \psi^*_{11}\psi^*_{21}\psi_{22}\psi_{12}
\mbox{~~and~~}
O^{z}_{21}\equiv \psi^*_{12}\psi^*_{22}\psi_{21}\psi_{11}
\label{s2zdops}
\end{equation}
for which (with the $\cO(1)$ constant $\xi\equiv
e^{\gamma_E}/2$)\footnote{Note that there is no arbitrariness here because
these composite operators do not require multiplicative renormalization for $m_f=0$
so the position-space correlators are well-defined for non-zero
separation.  A subtractive renormalization is required for the 2-point
function at zero separation.
}
\begin{equation}
\displaystyle
\braket{0|{\rm T}\,O^{z}_{12}(x)\,O^{z}_{21}(0)|0}
=\frac{\left(\xi m\right)^4}{16\pi^4}
\,\exp\left(4
K_0\left(m\sqrt{-x^2 + i\epsilon}\right) \right)
\label{2-2pt-11}
\end{equation}
These ``have zero anomalous dimension'' --- that is they go to constants at
long distances.  They 
were called ZDOPs (for Zero-Dimension OPerators) in
\cite{Georgi:2019tch} and I adopt that acronym here.
Cluster decomposition requires that these operators have VEVs,
\begin{equation}
\braket{0|O^{z}_{12}(0)|0}
=e^{i\theta}\frac{(\xi m)^2}{4\pi^2}
\quad\quad
\braket{0|O^{z}_{21}(0)|0}
=e^{-i\theta}\frac{(\xi m)^2}{4\pi^2}
\label{2-2pt-local-vev}
\end{equation}
where $\theta$ is a parameter that labels the vacuum
state~\cite{Smilga:1992hx,Jayewardena:1988td,Hetrick:1988yg}. 

\section{Conformal Coalescence\label{cc}}

For simplicity, I will focus on the dimension $1/2$ operators with zero
flavor $U(1)$ charge,
\be
O_1\equiv\psi^*_{1\,1}\psi_{1\,2}
\quad\quad
O_2\equiv \psi^*_{2\,1}\psi_{2\,2}
\ee
and their complex conjugates
\be
O^\cc_1=\psi^*_{1\,2}\psi_{1\,1}
\quad\quad
O^\cc_2=\psi^*_{2\,2}\psi_{2\,1}
\ee
The theory has a conserved axial isospin symmetry associated with the
charges
\begin{equation}
\vec I=\frac{1}{2}\int dx^1\,\bar\psi(x) \,\vec\sigma\,\gamma^5\,\psi(x) 
\label{axialcharges}
\end{equation}
where the Pauli matrices $\vec \sigma$ act on the flavor space
and the operators $O_1$ and $O_2$ have opposite charges for the 3rd component,
$I_3=+1$ and $-1$. 
The perturbative 2-pt functions are
\begin{equation}
\braket{0|T\,O_j(x)\,O^\cc_k(0)|0}
=\delta_{jk} \frac{\xi m}{(2\pi)^2}
\exp\left[K_0\left(m\sqrt{-x^2 + i\epsilon}\right) \right]
\left(\frac{1}{-x^2+i\epsilon}\right)^{1/2}\,
\label{s2-d1/2-p}
\end{equation}

The ZDOPs produce non-perturbative corrections to (\ref{s2-d1/2-p}).  
The perturbative 3-point correlation function
with an added ZDOP can be written as
{\renewcommand{\arraystretch}{2.5}
\begin{equation}
\begin{array}{c}
\displaystyle
\braket{0|T\,
O^{z}_{21}(z)
\,O_1(x)
O_2(0)|0}
\\ \displaystyle
=\braket{0|T\,
O^{z}_{12}(z)
\,O^\cc_1(x)
O^\cc_2(0)|0}
=
\frac{(\xi m)^3}{(2\pi)^4}
\left(\frac{1}{-x^2+i\epsilon}\right)^{1/2}
\\ \displaystyle
\exp\left[2\kappa_0(z-x)+2\kappa_0(z)-\kappa_0(x)\right]
\end{array}
\label{s2-d1/2-zdop}
\end{equation}}%
where
\be
\kappa_0(x)=K_0\left(m\sqrt{-x^2+i \epsilon}\right)
\ee

The form of (\ref{s2-d1/2-zdop}) can be understood (and indeed
overdetermined) as follows.  The overall 
counting of factors of $2\pi$ comes from the free fermion skeleton and is thus
the same as (\ref{2-2pt-11}).  The long-distance behavior is determined by
the anomalous dimension and because the ZDOP has zero anomalous dimension,
there is no long-distance dependence on $z$.  The $z$-dependence must be
entirely in the $K_0$ terms, which are determined by the gauge coupling and
which must combine to agree with (\ref{2-2pt-11}) as $x\to0$.  
The long-distance $x$ dependence and
thus the $1/(-x^2+i\epsilon)^{1/2}$ term must be
the same as in (\ref{s2-d1/2-p}). 
There is
no contribution to the $x$ dependence from the free fermion skeleton and
thus the $x$ dependence from the  $1/(-x^2+i\epsilon)^{1/2}$ term must cancel the
$x$ dependence from the $K_0$s at short distances, which fixes the
coefficient of $K_0\left(m\sqrt{-x^2 + i\epsilon}\right)$ in the
exponential.  The power of $(\xi m)$ is equal to sum of the coefficients of
the $K_0$s in the exponential.

Now cluster decomposition can 
be applied to (\ref{s2-d1/2-zdop}) just as it can in
(\ref{2-2pt-11}).  We can pull the ZDOP away
to infinity and replace it by its VEV, (\ref{2-2pt-local-vev}), then the
exponential in (\ref{s2-d1/2-zdop}) goes to $1$ and what remains is a nonperturbative
contribution to the 2pt functions of the dimension 1/2 operators. 
Thus\footnote{Note that if $x\to0$ in (\ref{s2-d1/2-np}), this reduces to
(\ref{2-2pt-local-vev}).} 
{\renewcommand{\arraystretch}{2.5}
\begin{equation}
\begin{array}{c}
\displaystyle
\braket{0|T\,
O_1(x)
O_2(0)|0}
=e^{-i\theta}
\frac{(\xi
m)}{(2\pi)^2}
\exp
\left[
-K_0\left(m\sqrt{-x^2 + i\epsilon}\right)\right]
\left(\frac{1}{-x^2+i\epsilon}\right)^{1/2}
\\ \displaystyle
\braket{0|T\,
O^\cc_1(x)
O^\cc_2(0)|0}
=e^{i\theta}
\frac{(\xi
m)}{(2\pi)^2}
\exp
\left[
-K_0\left(m\sqrt{-x^2 + i\epsilon}\right)\right]
\left(\frac{1}{-x^2+i\epsilon}\right)^{1/2}
\end{array}
\label{s2-d1/2-np}
\end{equation}}%
The ZDOP VEV has given us a
nonperturbative contribution to the 2-point function that 
is fixed by the calculable 3-point function.   It is
amusing that we can calculate this exactly.\footnote{In general, we might
have to include the contributions from $n$-point functions with more ZDOPs,
but in this example, these do not give any new contributions.}  But there
are more surprises 
in store.  Define 
\begin{equation}
O_g\equiv
e^{i\theta/2}O_1
+g\,
e^{-i\theta/2}O^\cc_2
\quad\quad
O^\cc_g\equiv
e^{-i\theta/2}O^\cc_1
+g\,
e^{i\theta/2}O_2
\label{og}
\end{equation}
where $g=\pm1$.

Then
\begin{equation}
\braket{0|T\,O_{\pm1}(x)\, O_{\pm1}(0)|0}
=\braket{0|T\,O_{\pm1}(x)\, O_{\mp1}(0)|0}
=\braket{0|T\,O_{\pm1}(x)\, O^\cc_{\mp1}(0)|0}
=0
\label{pmmp}
\end{equation}
The first two terms in (\ref{pmmp}) must vanish because of axial isospin
symmetry.  The vanishing of the third term follows because the
parameter $g$ is the multiplicative quantum number for a $\theta$-dependent 
G-parity that is conserved by 
perturbative and nonperturbative interactions.
\begin{equation}
e^{i\theta/2}O_1
\leftrightarrow
e^{-i\theta/2}O^\cc_2
\quad\quad
e^{i\theta/2}O_2
\leftrightarrow
e^{-i\theta/2}O^\cc_1
\label{gp}
\end{equation}

The only non-zero 2-point functions are
\begin{equation}
\begin{array}{c}
\displaystyle
\braket{0|T\,O_{\pm1}(x)\, O^\cc_{\pm1}(0)|0}=
\frac{(\xi m)}{4\pi^2}\times
\\ \displaystyle
2\left(\exp\left[K_0\left(m\sqrt{-x^2 + i\epsilon}\right)\right]\pm
\exp\left[-K_0\left(m\sqrt{-x^2 + i\epsilon}\right)\right]\right)
\left(\frac{1}{-x^2+i\epsilon}\right)^{1/2}
\end{array}
\label{pmpm}
\end{equation}
At short distances,
the first exponential in the penultimate factor in (\ref{pmpm})
dominates for both $+$ and $-$ and (along with the last
factor) produces the expected
free-fermion scaling.  But at long distances,
while the $O_{+1}$ operator goes smoothly to a conformal operator, the $O_{-1}$
correlator goes to zero exponentially.  One of the operators, the
$O_{-1}$, disappears 
from the conformal theory as the $O_1$ and $O^\cc_2$ pair in $O_{+1}$
coalesce!
Similar behavior was discovered in 1+1 diagonal color
models in \cite{Georgi:2019tch} and dubbed ``Conformal
coalescence.''  Here we will see that it has dramatic consequences in the
massive Schwinger model.

It is straightforward (if not particularly edifying) to write down the
general result.
{\renewcommand{\arraystretch}{3}
\begin{equation}
\begin{array}{c}
\displaystyle
\Braket{0|T\,\prod_{j=1}^n O_{g_j}(x_j)\,O^\cc_{h_j}(y_j)
|0}
=\left(\frac{\xi m}{4\pi^2}\right)^n
\\ \displaystyle
\left[
\frac
{
\prod_{j<k}
\biggl(\bigl(-(x_j-x_k)^2+i \epsilon\bigr)\bigl(-(y_j-y_k)^2+i\epsilon\bigr)\biggr)}
{\prod_{j,k}\bigl(-(x_j-y_k)^2+i \epsilon\bigr)}
\right]^{1/2}
\\ \displaystyle
\sum_{\eta_j,\chi_j=0}^1
\left(\prod_j(g_j)^{\eta_j+1}\right)
\left(\prod_j(h_j)^{\chi_j+1}\right)
\exp
\Biggl[
\left(\sum_{j,k}(-1)^{\eta_j+\chi_k}\kappa_0(x_j-y_k)\right)
\\ \displaystyle
-\left(\sum_{j<k}(-1)^{\eta_j+\eta_k}\kappa_0(x_j-x_k)
+(-1)^{\chi_j+\chi_k}\kappa_0(y_j-y_k)\right)
\Biggr]
\end{array}
\label{s2n}
\end{equation}
}
Again these vanish identically if the number of $O_{-1}$s plus the number of
$O^{\cc}_{-1}$s is odd, and vanish exponentially any of the 
$O_{-1}$ or
$O^{\cc}_{-1}$ coordinates goes to infinity.

\section{Mass terms\label{sec-mass}}

While I find this model endlessly fascinating, it is still just a generalized
free theory~\cite{Greenberg:1961mr}.  But I believe that the above analysis
can help us to understand the nontrivial theory that results from adding a
mass term.  This has been discussed in many works, but as I mentioned in
the introduction, I want to focus on the
three puzzles about the strong coupling limit, $m_f\ll m$ identified by
Coleman in \cite{Coleman:1976uz}. \\
{\it 
(1) Why are the lightest particles in the theory a degenerate isotriplet, even if
one quark is 10 times heavier than the other?\\
(2) Why does the next-lightest particle have $I^{PG}= 0^{++}$, rather than $0^{--}$?\\
(3) For $|\theta|=\pi$, how can an isodoublet quark and an isodoublet antiquark,
carrying opposite electric charges, make an isodoublet bound state with electric
charge zero?} 

I believe that conformal coalescence resolves the first puzzle in a
very simple way.  For $\theta=0$, (\ref{og}) implies that
an isospin invariant fermion mass term at low energies is
\be
m_f(O_1+O_2)+\mbox{h.c.}
=m_f(O_{+1}+O^\cc_{+1})
\to\frac{m_f\sqrt{\xi m}}{\pi}(O_{1/2}+O^\cc_{1/2})
\label{og0mf}
\ee
where $O_{1/2}$ is a normalized dimension $1/2$ conformal operator with
\begin{equation}
\braket{0|T\,O_{1/2}(x)\, O^\cc_{1/2}(0)|0}=
\left(\frac{1}{-x^2+i\epsilon}\right)^{1/2}
\label{o1/2}
\end{equation}
Note that (\ref{og0mf}) implies that the only quantity with dimensions that
survives in the low energy theory is $m_f\sqrt{m}$ and so the masses of the
particles that appear as a result of the breaking of the conformal symmetry
must be proportional to $(m_f^2m)^{1/3}$, in agreement with Coleman's
result.\footnote{It is easy to see that if $\delta m=0$ for arbitrary
$\theta\neq\pm\pi$, the mass parameter becomes
$m_f\sqrt{m}\cos\frac{\theta}{2}$, also in agreement with Coleman.}  

In the presence of an isospin breaking term for $\theta=0$
(\ref{og0mf}) goes to
\be
m_f(O_1+O_2)+\mbox{h.c.}
+\delta m(O_1-O_2)+\mbox{h.c.}
=m_f(O_{+1}+O^\cc_{+1})
+\delta m(O_{-1}+O^\cc_{-1})
\label{og0deltam}
\ee
All correlators involving the $\delta m$ term
go to zero exponentially at long distances because they involve powers of
$K_0(mx)$.  Because the only mass scale in the low energy theory is
$(m_f^2m)^{1/3}$, we expect that 
the isospin-breaking 
contribution is suppressed by powers of\footnote{The power of $2/3$ was
missing in an earlier version.  I am grateful to a referee for encouraging
me to make the argument more explicit. } 
\be
K_0\Bigl(m/(m_f^2m)^{1/3}\Bigr)
\propto \exp\Bigl(-(m/m_f)^{2/3}\Bigr)
\ee
even if $\delta m\approx m_f$.

I believe that the resolution of the second puzzle is in some sense obvious
but that it is telling us something novel about the conformal theory.  For
$\theta=0$, (\ref{og}) implies that the unparticle stuff produced by
$O_{+1}$ and $O^\cc_{+1}$ is G-even.  The G-odd stuff produced by $O_{-1}$
and $O^\cc_{-1}$ always involves the massive gauge boson and does not
survive at long distances!  Evidently, if we think of decreasing $m_f/m$
from weak coupling, $m_f/m\gg1$, to strong coupling, $m_f/m\ll1$, the G-odd
quark-antiquark
states get stuck at masses of order $m$ while the G-even states continue to
move down into the low energy theory.

Finally, I believe that the resolution of the third puzzle is that it is a
problem of logic rather than a problem of physics.  The puzzle starts from
the hypothesis that the low energy theory for an isospin invariant mass
term with $\theta=\pi$ is a theory of particles.  I believe that this
hypothesis is false.  For $\theta=\pi$, (\ref{og}) implies that the
isospin-invariant mass term
\be
m_f(O_1+O_2)+\mbox{h.c.}
=-i\,m_f(O_{-1}-O^\cc_{-1})
\to0
\label{ogpim}
\ee
Thus this term does not survive in the low energy conformal theory and the
low-energy conformal symmetry persists even in the presence of the mass
term.  If this is correct, then I think that 
there must be a phase transition between weak and strong coupling that frustrates
Coleman's attempt to understand the model in terms of the naive quark model.

\section{Directions for Future Work\label{sec-future}}

While I believe that I have answered each of Coleman's questions, the
answers suggest further questions.
For $\theta=0$ with non-zero $\delta m$, isospin symmetry breaking effects
are present at low energies,
but exponentially suppressed. In the low-energy theory, this like looks
fine-tuning.  Is this new mechanism for generating an exponential hierarchy
of parameters
useful for any of the hierarchy puzzles that afflict the standard model?
Is there a more physical description of what it means for the unparticle
stuff to have only even G-parity?  And for $\theta=\pi$, what does the
transition to the long-distance conformal theory look like?  I hope to
explore these questions further.  I wish that Sidney were still here to help!

\section*{Acknowledgements\label{sec-ack}}

I am grateful for insights from Bea Noether who was involved in the early
stages of this work.  I am also grateful for comments by Jacob Barandes,
Alvaro DeRujula,
David Kaplan, 
and Ashvin Vishwanath.
This work was supported in part by NSF grant  
PHY-1719924.  

\bibliography{up4}

\providecommand{\href}[2]{#2}\begingroup\raggedright\begin{thebibliography}{10}

\bibitem{Coleman:1976uz}
S.~R. Coleman, ``{More about the massive Schwinger model},''
\href{http://dx.doi.org/10.1016/0003-4916(76)90280-3}{{\em Ann. Phys.} {\bf
  101} (1976)  239}.

\bibitem{Georgi:2019tch}
H.~Georgi and B.~Noether, ``{Non-perturbative Effects and Unparticle Physics in
  Generalized Schwinger Models},'' \href{http://arxiv.org/abs/1908.03279}{{\tt
  arXiv:1908.03279 [hep-th]}}.

\bibitem{Belvedere:1978fj}
L.~V. Belvedere, K.~D. Rothe, B.~Schroer, and J.~A. Swieca, ``{Generalized
  Two-dimensional Abelian Gauge Theories and Confinement},''
\href{http://dx.doi.org/10.1016/0550-3213(79)90594-7}{{\em Nucl. Phys.} {\bf
  B153} (1979)  112--140}.

\bibitem{GamboaSaravi:1981zd}
R.~Gamboa~Saravi, F.~Schaposnik, and J.~Solomin, ``{Path Integral Formulation
  of Two-dimensional Gauge Theories With Massless Fermions},''
  \href{http://dx.doi.org/10.1016/0550-3213(81)90375-8}{{\em Nucl. Phys. B}
  {\bf 185} (1981)  239--253}.

\bibitem{Gattringer:1993ec}
C.~Gattringer and E.~Seiler, ``{Functional integral approach to the N flavor
  Schwinger model},'' \href{http://dx.doi.org/10.1006/aphy.1994.1062}{{\em
  Annals Phys.} {\bf 233} (1994)  97--124},
\href{http://arxiv.org/abs/hep-th/9312102}{{\tt arXiv:hep-th/9312102
  [hep-th]}}.

\bibitem{Delphenich:1997ex}
D.~Delphenich and J.~Schechter, ``{Multiflavor massive Schwinger model with
  nonAbelian bosonization},''
  \href{http://dx.doi.org/10.1142/S0217751X9700284X}{{\em Int. J. Mod. Phys. A}
  {\bf 12} (1997)  5305--5324}, \href{http://arxiv.org/abs/hep-th/9703120}{{\tt
  arXiv:hep-th/9703120}}.

\bibitem{Nagy:2008ys}
S.~Nagy, ``{Massless fermions in multi-flavor QED(2)},''
  \href{http://dx.doi.org/10.1103/PhysRevD.79.045004}{{\em Phys. Rev. D} {\bf
  79} (2009)  045004}, \href{http://arxiv.org/abs/0805.2009}{{\tt
  arXiv:0805.2009 [hep-th]}}.

\bibitem{Azcoiti:2019moz}
V.~Azcoiti, ``{Interplay between $SU(N_f)$ chiral symmetry, $U(1)_A$ axial
  anomaly, and massless bosons},''
  \href{http://dx.doi.org/10.1103/PhysRevD.100.074511}{{\em Phys. Rev. D} {\bf
  100} (2019) no.~7, 074511}, \href{http://arxiv.org/abs/1907.01872}{{\tt
  arXiv:1907.01872 [hep-lat]}}.

\bibitem{Banks:1981nn}
T.~Banks and A.~Zaks, ``{On the phase structure of vector-like gauge theories
  with massless fermions},''
\href{http://dx.doi.org/10.1016/0550-3213(82)90035-9}{{\em Nucl. Phys.} {\bf
  B196} (1982)  189}.

\bibitem{Smilga:1992hx}
A.~V. Smilga, ``{On the fermion condensate in the Schwinger model},''
\href{http://dx.doi.org/10.1016/0370-2693(92)90209-M}{{\em Phys. Lett.} {\bf
  B278} (1992)  371}.

\bibitem{Jayewardena:1988td}
C.~Jayewardena, ``{SCHWINGER MODEL ON S(2)},''
{\em Helv. Phys. Acta} {\bf 61} (1988)  636--711.

\bibitem{Hetrick:1988yg}
J.~E. Hetrick and Y.~Hosotani, ``{QED ON A CIRCLE},''
\href{http://dx.doi.org/10.1103/PhysRevD.38.2621}{{\em Phys. Rev.} {\bf D38}
  (1988)  2621}.

\bibitem{Greenberg:1961mr}
O.~W. Greenberg, ``{Generalized Free Fields and Models of Local Field
  Theory},''
\href{http://dx.doi.org/10.1016/0003-4916(61)90032-X}{{\em Annals Phys.} {\bf
  16} (1961)  158--176}.

\end{thebibliography}\endgroup

\end{document}